\documentclass[preprint2]{aastex}
\usepackage[utf8]{inputenc}
\usepackage{multirow}
\usepackage{rotating}
\usepackage{longtable}
\usepackage{lscape}
\usepackage{hhline}
\slugcomment{To appear in ApJ Main Journal}
\shorttitle{Dust in the Magellanic Clouds}
\shortauthors{A. Zonca et al.}

\newcommand{\noprint}[1]{}

\begin{document}
\title{Modelling dust extinction in the Magellanic Clouds}
\author{Alberto Zonca, Silvia Casu, Giacomo Mulas, Giambattista Aresu}
\affil{INAF\textendash Osservatorio Astronomico di Cagliari, Via della Scienza 5, I-09047 Selargius, Italy}
\email{azonca@oa-cagliari.inaf.it,silvia@oa-cagliari.inaf.it,gmulas@oa-cagliari.inaf.it, \\ garesu@oa-cagliari.inaf.it}
\author{Cesare Cecchi\textendash Pestellini}
\affil{INAF\textendash Osservatorio Astronomico di Palermo, P.za Parlamento 1, I-90134 Palermo, Italy} 
\email{cecchi-pestellini@astropa.inaf.it}

\date{Accepted. Received; in original form}
\begin{abstract}
We model the extinction profiles observed in the Small and Large Magellanic clouds with a synthetic population of dust grains consisting by core-mantle particles and a collection of free\textendash flying polycyclic aromatic hydrocarbons. All different flavors of the extinction curves observed in the Magellanic Clouds can be described by the present model, that has been previously (successfully) applied to a large sample of diffuse and translucent lines of sight in the Milky Way. We find that in the Magellanic Clouds the extinction produced by classic grains is generally larger than absorption by polycyclic aromatic hydrocarbons. Within this model, the non\textendash linear far\textendash UV rise is accounted for by polycyclic aromatic hydrocarbons, whose presence in turn is always associated to a gap in the size distribution of classical particles. This hints either a physical connection between (e.g., a common cause for) polycyclic aromatic hydrocarbons and the absence of middle\textendash sized dust particles, or the need for an additional component in the model, that can account for the non\textendash linear far\textendash UV rise without contributing to the UV bump at $\sim$217~nm, e.g., nanodiamonds.
\end{abstract}

\section{Introduction}\label{introduction}
InterStellar Extinction Curves (ISECs) are now available for many lines of sight in the Milky Way Galaxy (MWG), in the Small and Large Magellanic Clouds  (SMC, LMC), in M31, in nearby galaxies, and increasingly in objects at high redshift. The ISECs have a great variety of shapes. This evidence may seem to suggest  that grain models must be complex, allowing for different carriers for each of the different components of the ISECs. However, all the measured extinction profiles are clearly members of the same family, and in fact, ISECs can be phenomenologically described by means of a very limited number of empirical parameters \citep{CCM,V04,FM07}. Such a smooth topological transition among extinction profiles may be naturally interpreted as the response of dust to the local physics. 

Dust grains are composed mainly of silicates and carbons, as made clear by spectroscopic evidence and depletion data (e.g., \citealt{D03}). They are formed in the envelopes of cool stars and in novae and supernovae, and are  destroyed in shocks, but with different efficiencies, silicates being more resistant than carbons \citep{Z13,B14}. Even if the silicates and carbons can be produced as separate populations, the destruction of solid carbons to atoms and small molecules will lead to carbon deposition on silicate cores. According to the scenario originally envisaged by \citet{J90}, gaseous carbon in diffuse clouds is gradually deposited on silicate cores in hydrogen rich $sp^3$ form, and is eventually annealed and darkened to graphitic (H-poor) $sp^2$ kind by the interstellar radiation field. In other words, dust is not immutable and  is expected to evolve in time (e.g., \citealt{CCP10}). 

Such dust formation and evolution scenario has been implemented in a model ~\textendash~ the [CM]$^2$ model \citep{CCP08} ~\textendash~ successfully applied to a sample of 329 galactic lines of sights \citep{FM07} by \citet{M13}.   

Subsequently, the carbon evolution prescriptions of this scenario were condensed in a set of time\textendash dependent equations, describing carbon deposition onto the silicate cores and its following photo-darkening,  and matched to the observationally based inferences of dust physical parameters \citep{CCP14}, closing the circle. Most of the ISECs in the \citet{FM07} sample correspond to dust mantles that restarted growing after being almost completely removed, with the leftover fully rehydrogenated, by, e.g., occasional shocks. 

A natural extension of this work is its application to (and test with) environments markedly different from the interstellar medium of the solar neighborhood, in which the life cycle of the interstellar medium is known to be very different either globally, e.g., much more frequent shocks as in starburst galaxies, or locally, because of independent observational constraints on individual lines of sight. In this paper we shall apply the [CM]$^2$ model to a number of lines of sight in the Magellanic Clouds (MCs), whose data have been provided by \citet{G03} and \citet{C05}. These two irregular dwarf galaxies are nearby ($\sim 50$~kpc), gas\textendash rich, actively star-forming, and have sub-solar metallicities (e.g., \citealt{Howarth2011}). The SMC average ISEC presents a featureless profile with no 217.5 nm bump and a steep far ultraviolet rise, while the LMC one shows a hint of the bump and a firm far-ultraviolet rise, stronger than in the MWG but less sharp than in the SMC.

We describe the specific lines of sight, their known physical properties, and how they are modelled in Section~\ref{MOD}, the resulting characterisation of dust grains along such extinction paths in Section~\ref{RES}, Section~\ref{discussection} contains our discussion and the last Section summarises our conclusions.

\section{Extinction observed and modelled}
\label{MOD}
 \citet{G03} and \citet{C05} exploited a collection of newly taken and archive data to obtain individual extinction curves for a total of 24 lines of sight: 19 for LMC and 5 for SMC. The list of sight lines reported in Table~\ref{one} contains also their absolute visual magnitude, $A_V$, the ratio of total to selective extinction, $R_V$, and the gas to dust ratio expressed as $N_{\rm H}/E_{B-V}$ cm$^{-2}$~mag$^{-1}$. This quantity has been constructed for specific lines of sight using data taken from \citet{W12} and references therein: the atomic hydrogen column density is derived from absorption Lyman-$\alpha$ profiles in all cases but Sk-69~279, for which we used data from the hydrogen 21 cm emission; the molecular hydrogen column density is obtained  from \emph{FUSE} H$_2$ absorption measurements when available, or otherwise exploiting  the relation linking $N_{\rm H_2}/E_{B-V}$ to $E_{B-V}$, as found by \citet{W12} for the MWG. In addition to individual lines of sight, we also considered the ``average'' extinction curves given by \citet{G03} for the SMC bar, the LMC~2 Supershell (Ss), and the overall LMC average. These average extinction curves were treated exactly as the individual ones, to see if this resulted in a sensible description of representative dust properties for the areas they refer to.
\begin{table*}
\centering
\caption{Relevant properties of the lines of sight toward the Magellanic Clouds in our sample (in italic the average ISECs).}
\begin{tabular}{lccc}
\\
\hline \hline
LoS & $A_V$ & $R_V$ & $N_{\rm H}/E_{B-V}$ \\
& (mag) & &  ($\times 10^{21}$ cm$^{-2}$~mag$^{-1}$) \\
\hline
\multicolumn{4}{c}{\footnotesize SMC bar sample} \\
AzV18    & 0.49$^1$ & 2.90$^1$ & 67.19$^{3,4}$ \\  
AzV23    & 0.48$^2$ & 2.65$^2$ & 49.29$^{3,5}$  \\
AzV214  & 0.35$^2$ & 2.40$^2$ & 41.03$^{3,5}$  \\  
AzV398  & 0.68$^2$ & 3.14$^2$ & 55.63$^{3,5}$  \\
\emph{SMC Bar} & \textemdash & 2.74$^2$ & 24.55$^{7}$  \\
\hline
\multicolumn{4}{c}{\footnotesize SMC wing sample} \\
AzV456  & 0.57$^1$ & 2.19$^1$ & 10.39$^{3,4}$ \\  
\hline
\hline
\multicolumn{4}{c}{\footnotesize LMC Average Sample} \\
Sk-66 19    & 0.86$^2$  & 3.44$^2$ & 29.59$^{3,4}$  \\ 
Sk-66 88    & 1.03$^2$  & 3.67$^2$ & 20.71$^{3,5}$  \\  
Sk-67 2      & 0.56$^1$  & 3.75$^1$ & 18.55$^{3,4}$  \\ 
Sk-68 23    & 1.04$^2$  & 3.35$^2$ &  6.03$^{3,5}$  \\ 
Sk-68 26    & 0.64$^1$  & 3.45$^1$ & 20.77$^{3,4}$  \\ 
Sk-68 129  & 0.57$^1$  & 3.37$^1$ & 32.73$^{3,4}$  \\ 
Sk-69 108  & 0.98$^2$  & 3.15$^2$ & 10.88$^{3,5}$   \\ 
Sk-69 206  & 0.96$^2$  & 3.68$^2$ & 35.33$^{3,5}$   \\ 
Sk-69 210  & 1.36$^2$  & 3.32$^2$ & 25.75$^{3,5}$   \\ 
Sk-69 213  & 0.63$^2$  & 3.96$^2$ & 22.55$^{3,5}$   \\ 
\emph{LMC} & \textemdash & 3.41$^2$ & 12.88$^{7}$   \\
\hline
\multicolumn{4}{c}{\footnotesize LMC2 Ss sample} \\
Sk-68 140  & 0.67$^1$  & 3.34$^1$ & 28.77$^{3,4}$  \\ 
Sk-68 155  & 0.56$^1$  & 2.81$^1$ & 21.82$^{3,4}$  \\
Sk-69 228  & 0.53$^1$  & 3.54$^1$ & 28.50$^{3,4}$   \\ 
Sk-69 256  & 0.11$^2$  & 0.64$^2$ & 15.67$^{3,5}$    \\ 
Sk-69 265  & 0.32$^2$  & 1.68$^2$ & 24.43$^{3,5}$   \\ 
Sk-69 270  & 0.44$^2$  & 2.34$^2$ & 18.99$^{3,4}$   \\ 
Sk-69 279  & 0.75$^1$  & 3.54$^1$ & 21.80$^{6,4}$   \\ 
Sk-69 280  & 0.56$^2$  & 3.12$^2$ & 23.03$^{3,5}$    \\ 
Sk-70 116  & 0.65$^2$  & 3.41$^2$ & 16.82$^{3,5}$    \\ 
\emph{LMC2 Ss}  & \textemdash & 2.76$^2$ & 21.88$^{7}$        \\
\hline \hline
\end{tabular}
\flushleft
$^1$ \cite{C05}; $^2$ \cite{G03}; $^3$ atomic hydrogen from Lyman $\alpha$; $^4$ molecular hydrogen from \emph{FUSE}; $^5$ molecular hydrogen estimate (see text); 
$^6$  atomic hydrogen from 21 cm; $^7$ \cite{W12}
\label{one}
\end{table*}

The dust model consists in a distribution of core-mantle grains and a mixture of polycyclic aromatic hydrocarbons (PAHs). Classical grains are constructed assembling four concentric components: a cavity, whose volume is in fixed proportion $f_v$ for all core sizes; the silicate shell of radius $a$; the $sp^2$ layer of thickness $f_{sp^2} w$, $w$ being the mantle thickness; and the $sp^3$ layer of thickness $f_{sp^3} w = (1- f_{sp^2}) w$. We consider a size distribution for the silicate cores, given by a power law $(a+w)^{-q}$. Such distribution allows for a gap in particle sizes, so that two populations of dust grains, big and small ones, may be present, each characterised by lower ($a_-$, $b_-$) and upper size ($a_+$, $b_+$) limits in the distribution. The molecular component is represented by a mixture of 54 PAHs in the size range $10 - 66$ C atoms, in the charge states 0, $\pm 1$, and +2 \citep{M07}. The equations describing the model have been reported in several papers (e.g., \citealt{CCP08,Z11}), and are not repeated here. The model produces a theoretical extinction curve as a function of its parameters. A least\textendash squares procedure \citep{levenberg1944, markwardt2009} yields the best\textendash fitting model parameter values for each line of sight. The variance\textendash covariance matrix for the best\textendash fitting parameters, measuring how tightly they are constrained by the fit, is obtained by synthetic statistics, by repeating the fit multiple times on data perturbed according to their errors. The details of this procedure, and the modifications in the MPFIT\footnote{http://purl.com/net/mpfit} Levenberg\textendash Marquardt implementation to cope with underdetermination of some parameters, are described in \citet{M13}.

The evolutionary model of carbon mantles is detailed in \citet{CCP14}. The carbon cycle in and out of dust is described by a set of ordinary differential equations that follow the evolution in time of the available gas\textendash phase carbon, the deposited polymeric $sp^3$ carbon , and its radiation-annealed $sp^2$ counterpart. The main free parameters are the photo\textendash darkening time ~\textemdash~ the time scale for graphitization (e.g., \citealt{CCP10}) ~\textemdash~and the initial conditions. The differential equations thus constitute a one\textendash parameter system: solutions starting from the same initial conditions, but with different photo\textendash darkening times, have no other points in common but the starting point and the asymptotic limit. Of course, the evolutionary model also has a parametric dependence on the assumed environmental conditions, i.e. the elemental abundances and so on. However, this dependence is rather mild, and the solutions of the model retain their overall behaviour with minor differences for any sensible environmental conditions \citep{CCP14}.

\section{Results}
\label{RES}
\begin{figure}
\begin{center}
\includegraphics[width=\hsize]{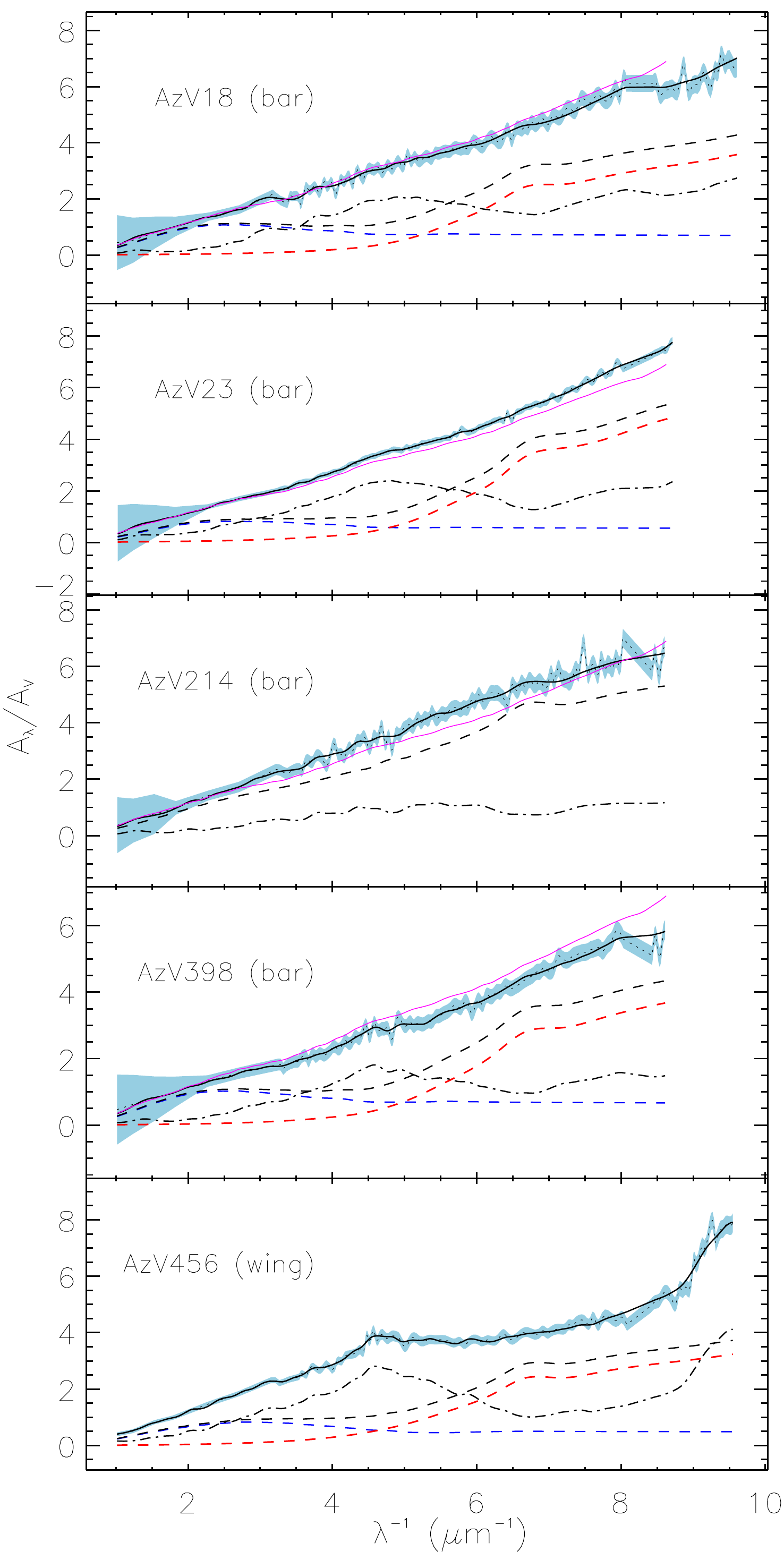} 
\caption{Fits to the normalized extinction curves for five lines of sight in the Small Magellanic Cloud. Solid black lines: global fit. Dashed lines: classical dust contribution; in blue the contribution of ``large'' dust grains, in red ``small'' dust grains, in black their sum; only the black curve is shown when the gap is very small or nonexistent. Dot\textendash dashed lines: PAHs. For lines of sight in the SMC bar, we also show, in solid purple, the ``average'' extinction curve for the SMC bar \citep{G03}. The lightly shaded areas are the observational error ranges (see text).}
\label{smcfitcurves}
\end{center}
\end{figure}

\begin{figure*}
\begin{tabular}{cc}
\includegraphics[scale=0.5]{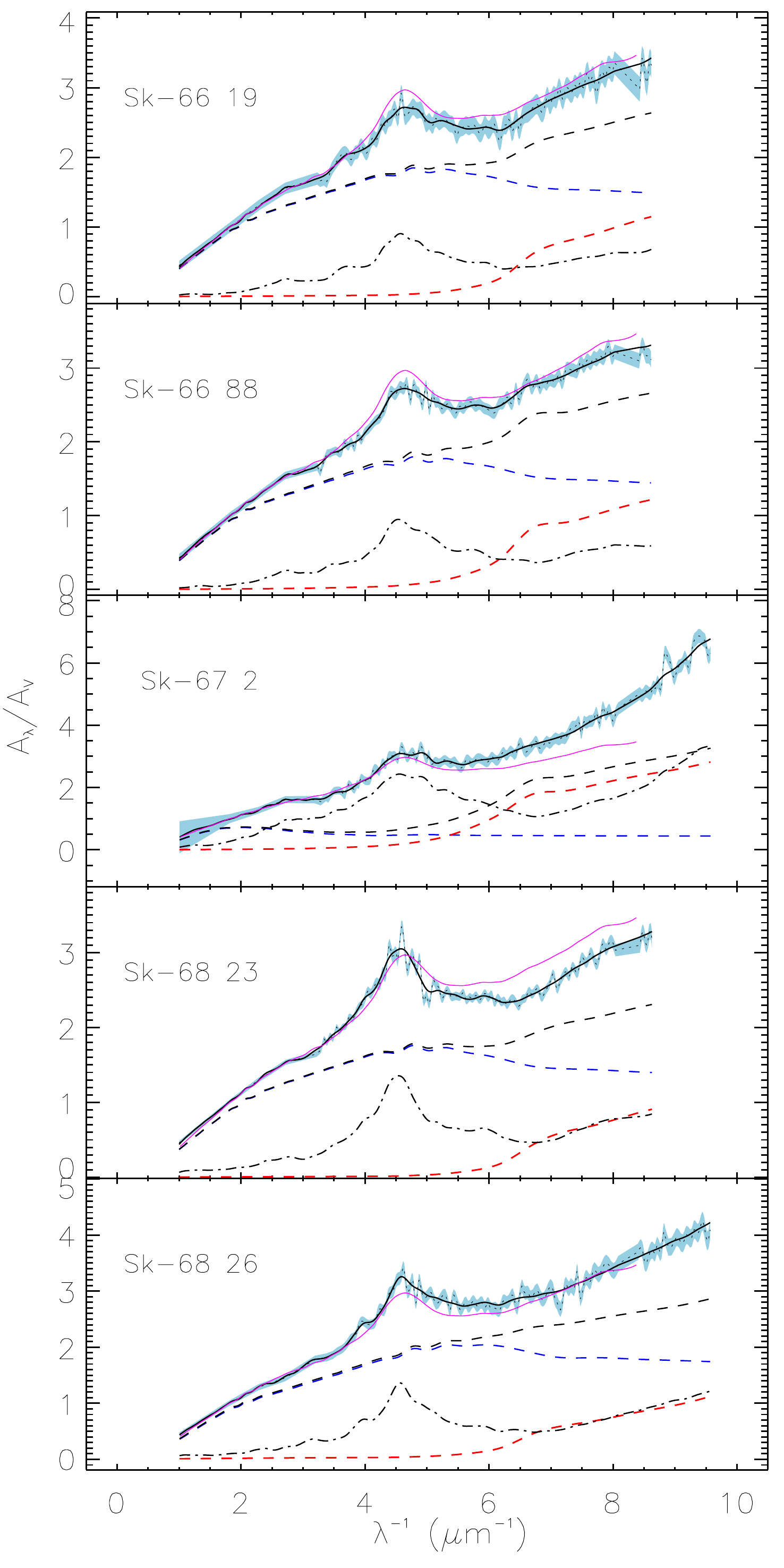} & \includegraphics[scale=0.5]{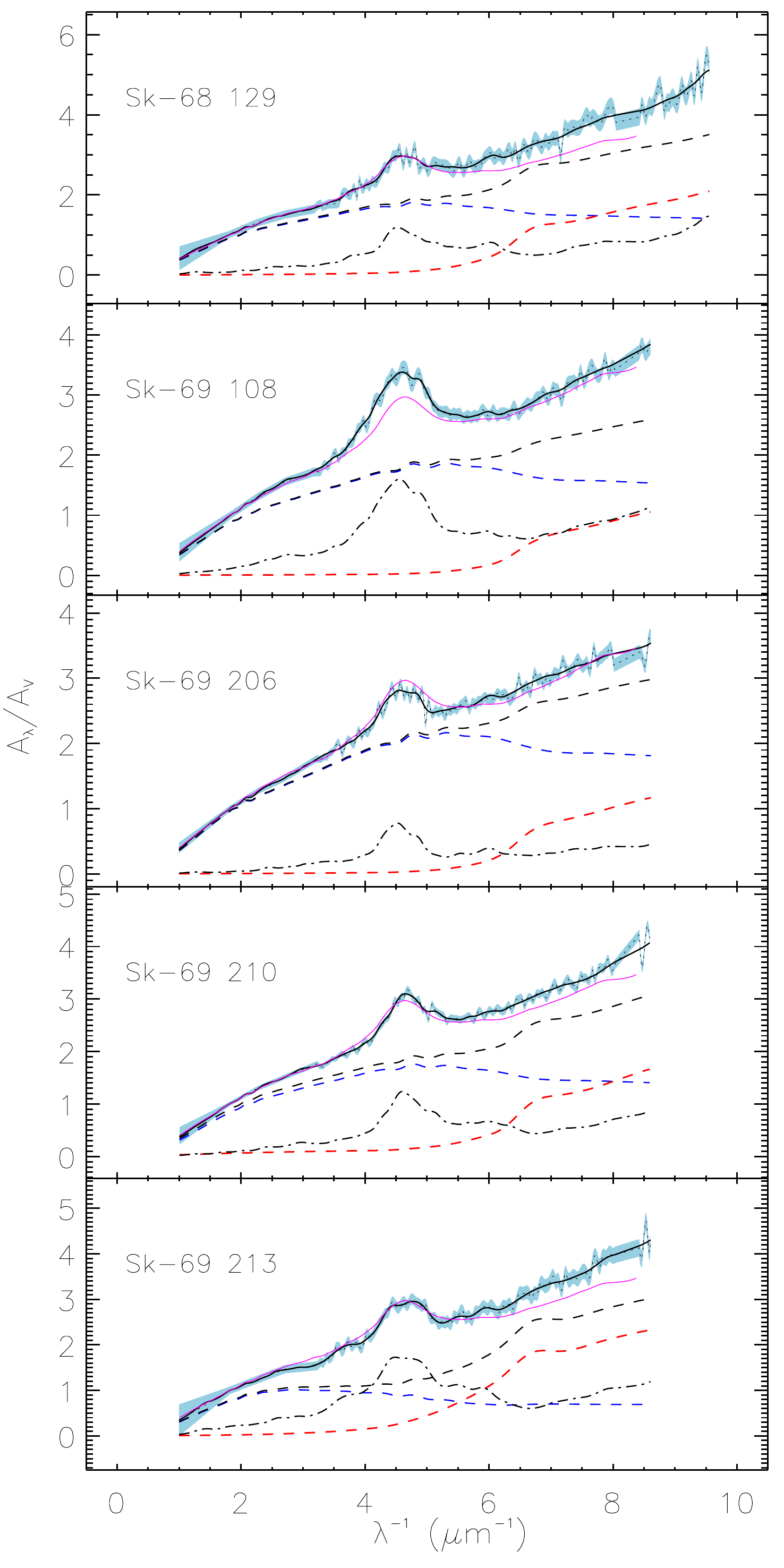}
\end{tabular}
\caption{Fits to the normalized extinction curves for lines of sight in the Large Magellanic Cloud. Solid black lines: global fit. Dashed lines: classical dust contribution; in blue the contribution of ``large'' dust grains, in red ``small'' dust grains, in black their sum; only the black curve is shown when the gap is very small or nonexistent. Dot\textendash dashed lines: PAHs. In solid purple, the ``average'' extinction curve for the LMC \citep{G03}.; dot\textendash dashed lines: PAHs. The lightly shaded areas are the observational error ranges (see text).}
\label{lmcfitcurves}
\end{figure*}

Fitted ISECs are shown in Figs.~\ref{smcfitcurves} (SMC) and \ref{lmcfitcurves} (LMC), while the inferred dust parameters are reported in Tables~\ref{classictable} (classical dust) and
\ref{pahtable} (PAHs). In the same tables we also show the results for the average extinction profiles.
\setcounter{figure}{1}
\begin{figure*}
\begin{tabular}{cc}
\includegraphics[scale=0.5]{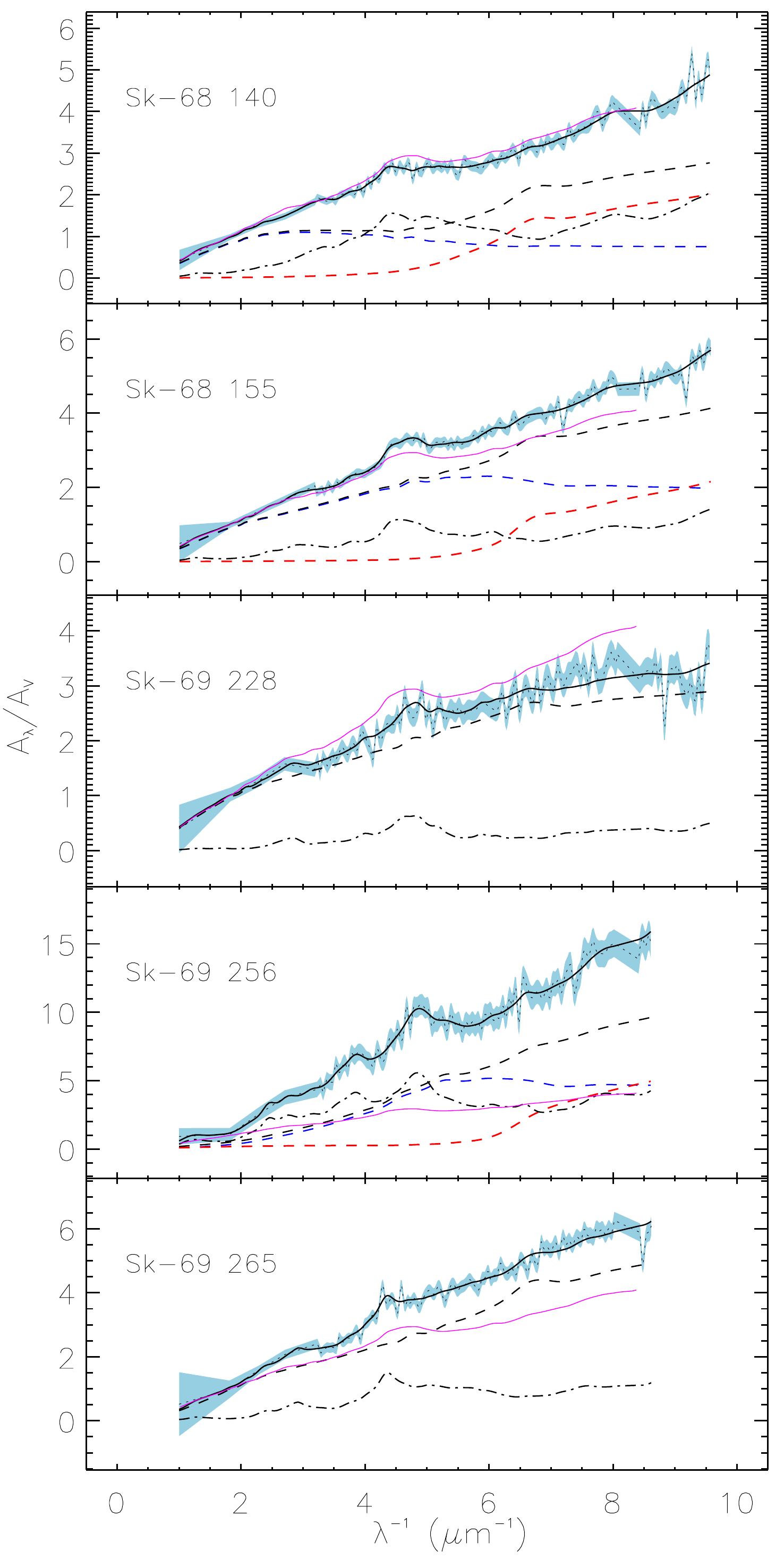} & \includegraphics[scale=0.5]{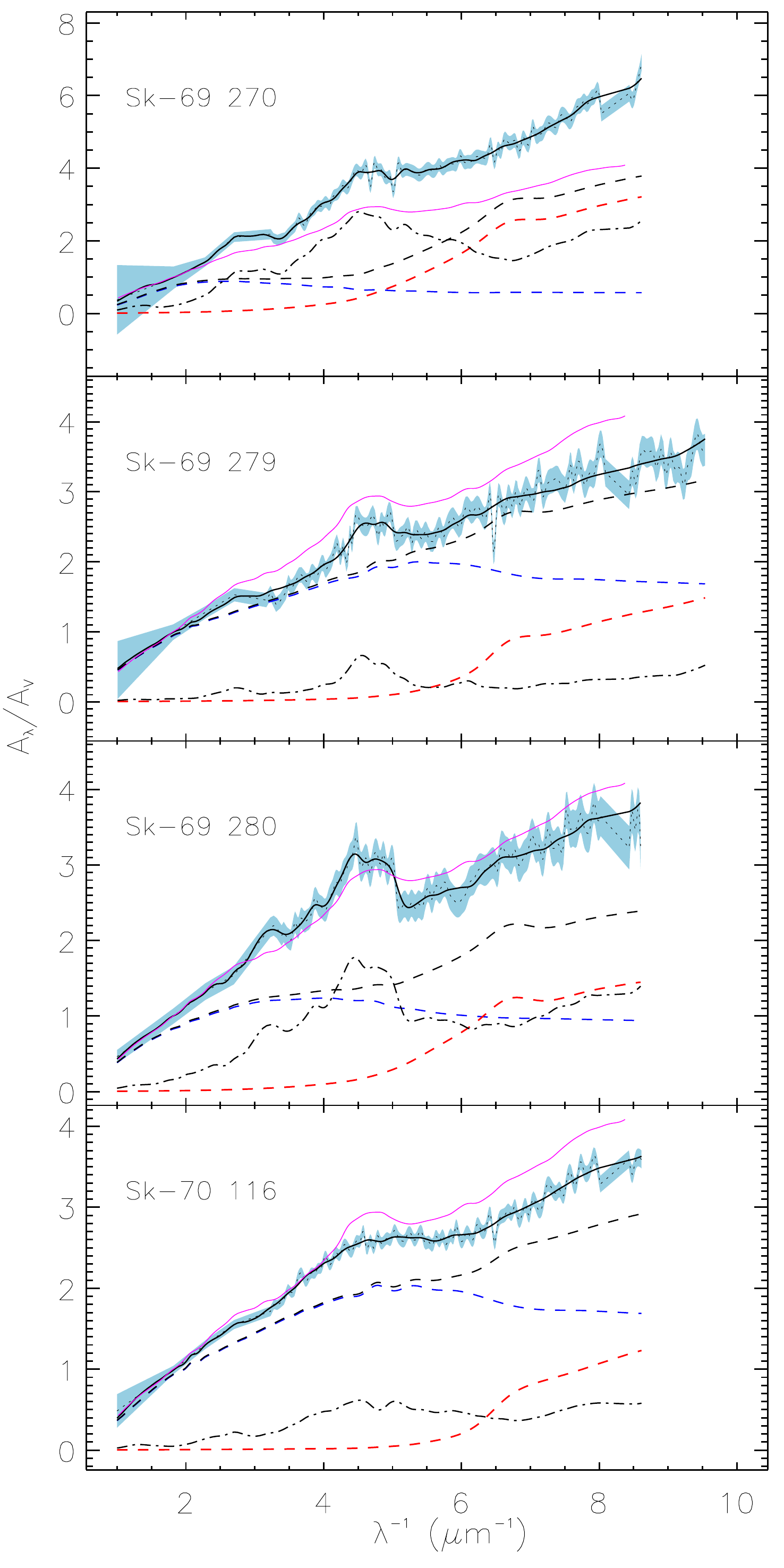}
\end{tabular}
\caption{ Fits to the normalized extinction curves for lines of sight in the Large Magellanic Cloud LMC2 Ss. Solid black lines: global fit. Dashed lines: classical dust contribution; in blue the contribution of ``large'' dust grains, in red ``small'' dust grains, in black their sum; only the black curve is shown when the gap is very small or nonexistent Dot\textendash dashed lines: PAHs. In solid purple, the ``average'' extinction curve for the LMC \citep{G03}.; dot\textendash dashed lines: PAHs. The lightly shaded areas are the observational error ranges (see text).}
\end{figure*}
{
\scriptsize
\begin{sidewaystable*}
\centering
\caption{Classical dust component parameters and Carbon column density locked in PAHs for each line of sight. }
\begin{scriptsize}
\begin{tabular}{cccccccccccc}
\hline
\hline 
$Los$ & $N_{d}$ & $f_{v}$ & $w$ & $f_{sp^{2}}$ & $a_{-}$ & $a_{+}$ & $b_{-}$ & $b_{+}$ & $\delta a$ ($ b_{-} - a_{+}$) & $q$ &  $N_{\rm C}^{\rm {PAH}} $ \tabularnewline
&&&&&&&&&&& $\times 10^{15}$(cm$^{-2}$) \tabularnewline
\hline
\hline
\\ \multicolumn{12}{c}{\footnotesize SMC bar sample} \\
\\
AzV18 & 1.80 	 (0.14) & 	0.00 	 ($<\varepsilon_{2}$)  & 	2.0 	 (0.9) & 	0.00	 ($<\varepsilon_{2}$) & 	7 	 (3) & 	42 	 (3)  & 	130 	 (18)  & 	282 	 (110) & 87 	 (19)
3.51 	 (0.04) &  137 	 (18)\\
AzV23$^{(a)}$ & 3.336 (0.003) & 0.67 ($<\varepsilon_{2}$) & 4.70 (0.02) & 0.77 ($<\varepsilon_{2}$) & 5.00  ($<\varepsilon_{2}$) & 136.0 (0.7) & \textemdash & \textemdash & 0.0 ($<\varepsilon_{2}$) & 3.3199 (0.0007) & 0.0     (0.0) \\ 
AzV23$^{(b)}$ &	3.87 	 (0.13) & 	0.00 	 ($<\varepsilon_{2}$) & 	0.3 	 (0.5) &	0.6 	 (0.4) &	5.07 	 (0.22) & 	45.7 	 (2.6) & 	118 	 (17)  & 	339 	 (200) & 72 	 (18) &	3.760 	 (0.016) & 178 	 10	\\
AzV214  & 	1.79 	 (0.12) &	0.55 	 (0.06)  & 	0.64 	 (0.14)  & 	0.00 	 ($<\varepsilon_{2}$)  & 	5.00 	 ($<\varepsilon_{2}$)  & 	607 	 (240)  & 	\textemdash  & 	\textemdash  &  	0.0 	 ($<\varepsilon_{2}$) &	3.527 	 (0.015) & 68 	 (13)    \\ 
AzV398 &	1.83 	 (0.12)  & 	0.01 	 (0.02)  & 	1.7 	 (0.7)  & 	0.00 	 ($<\varepsilon_{2}$)  & 	5.1 	 (0.4)  & 	44.0 	 (2.3)   & 	153 	 (50)  & 	345 	 (160)  & 109 	 (50) &	3.52 	 0.03 & 179 	 21\\ 
\emph{SMC Bar} & 2.10 	 (0.06) &	0.00 	 ($<\varepsilon_{2}$) & 	0.6 	 (0.3) & 	0.00 	 ($<\varepsilon_{2}$)   &	5.00 	 ($<\varepsilon_{2}$) &	44 	 (4) &	130 	 (8) &	269 	 (16) & 86 	 (10) &	3.583 	 (0.016) & \textemdash \\
\\ \hline
\\
 \multicolumn{12}{c}{\footnotesize SMC wing sample} \\
\\
AzV456 & 1.21 (0.11) & 0.39 (0.05) & 1.5 (0.7) & 0.00 ($<\varepsilon_{2}$) & 6.2 (1.6) & 66 (4) & 195 (26) & 410 (15) & 129 	 (28) & 3.458 (0.021) & 251 	 (11)   \\  
\\ \hline
\hline
\\
\multicolumn{12}{c}{\footnotesize LMC Average sample} \\
\\
Sk-66 19 & 1.00 (0.04) & 0.24 (0.03) & 1.5 (0.6) & 0.00 ($<\varepsilon_{2}$) & 5.00 ($<\varepsilon_{2}$) & 26 (4) & 60 (8) & 530 (30) & 34 	 (5) & 3.489 (0.007) & (97) 	 (14) \\ 
Sk-66 88 & 1.026 (0.022) & 0.29 (0.02) & 2.7 (0.4) & 0.00 ($<\varepsilon_{2}$) & 5.4 (1.5) & 34 (5) & 62 (15) & 542 (29) & 28 	 (10) & 3.487 (0.007) & 109 	 (19) \\ 
Sk-67 2 & 1.0706 (0.0027) & 0.00 ($<\varepsilon_{2}$) & 1.09 (0.05) & 0.00 ($<\varepsilon_{2}$) & 5.00 ($<\varepsilon_{2}$) & 37.54 (0.01) & 152.6 (0.4) & 337 (12) & 115.1 	 (0.4) & 3.4891 ($<\varepsilon_{3}$) & 192.3 	 (2.5) \\ 
Sk-68 23 & 0.966 (0.008) & 0.26 (0.01) & 1.1 (0.4) & 0.00 ($<\varepsilon_{2}$) & 5.00 ($<\varepsilon_{2}$) & 19.2 (1.9) & 62.0 (2.6) & 538 (9) & 42.8 	 (1.8) & 3.4896 (0.0021) &  145 	 (8)\\ 
Sk-68 26 & 0.966 (0.029) & 0.40 (0.03) & 1.7 (0.5) & 0.27 (0.17) & 5.00 ($<\varepsilon_{2}$) & 24 (4) & 47.9 (1.8) & 610 (16) & 24 	 (6) & 3.479 (0.004) & 88 	 (7)  \\
Sk-68 129 & 1.00 (0.08) & 0.29 (0.09) & 1.5 (0.9) & 0.00 ($<\varepsilon_{2}$) & 5.00 ($<\varepsilon_{2}$) & 43 (10) & 83 (26) & 544 (80) & 40 	 (17) & 3.481 (0.015) & 89 	 (19)  \\ 
Sk-69 108 & 0.97 (0.04) & 0.34 (0.04) & 1.3 (0.4) & 0.00 (0.01) & 5.05 (0.21) & 26 (4) & 59 (6) & 539 (50) & 34 	 (3) & 3.480 (0.006) & 168 	 (22)  \\ 
Sk-69 206 & 1.093 (0.021) & 0.48 (0.02) & 1.64 (0.27) & 0.00 ($<\varepsilon_{2}$) & 5.00 ($<\varepsilon_{2}$) & 27.3 (2.4) & 64.4 (2.0) & 589 (40) & 37.2 	 (2.4) & 3.452 (0.005) & 72 	 (9)  \\ 
Sk-69 210 & 1.038 (0.009) & 0.43 (0.01) & 0.56 (0.09) & 0.58 (0.09) & 5.00 ($<\varepsilon_{2}$) & 33.3 (1.1) & 74.4 (1.6) & 545 (40) & 41.1 	 (1.6) & 3.474 (0.004) & 185 	 (12) \\
Sk-69 213 & 1.01 (0.05) & 0.25 (0.09) & 1.15 (0.25) & 0.00 ($<\varepsilon_{2}$) & 5.00 ($<\varepsilon_{2}$) & 53 (4) & 137 (16) & 485 (60) & 84 	 (16) & 3.492 (0.009) & 153 	 (16) \\ 
\emph{LMC} & 1.14 	 (0.20) & 	0.25 	 (0.13)  & 	3.5 	 (1.5) &	0.01 	 (0.02) &	10 	 (6) &	47 	 (6)  & 	158 	 (70) & 	435 	 (60) &  112 	 (70) &	3.46 	 (0.04) &  \textemdash \\ 
\\ \hline
\\
\multicolumn{12}{c}{\footnotesize LMC2 Ss sample} \\
\\

Sk-68 140 & 1.08 (0.11) & 0.16 (0.09) & 3.2 (1.1) & 0.00 ($<\varepsilon_{2}$) & 5.5 (1.2) & 45 (4) & 122 (13) & 413 (90) & 77 	 (13) & 3.473 (0.020) & 146 	 (12)  \\ 
Sk-68 155 & 1.12 (0.05) & 0.39 (0.08) & 0.88 (0.15) & 0.00 ($<\varepsilon_{2}$) & 5.00 ($<\varepsilon_{2}$) & 40 (5) & 51 (3) & 684 (210) & 11 	 (5) &  3.493 (0.014) & 80 	 (9)   \\ 
Sk-69 228 & 0.910 (0.025) & 0.25 (0.08) & 4.7 (0.3) & 0.00 ($<\varepsilon_{2}$) & 5.2 (0.6) & 35.1 	 (1.5)  & 36.9 	 (1.9)  & 574 (90) & 1.3 	 (1.1) & 3.461 (0.008) & 28 	 (3)  \\ 
Sk-69 256$^{(c)}$ & 15.62 (0.12) & 0.67 ($<\varepsilon_{2}$) & 1.7 (1.3) & 0.27 (0.22) & 7 (5) & 13 (8) & 96 (22) & 109 (26) & 82 	 (21)  &  3.28 (0.03) & 104 	 (20) \\ 
Sk-69 256$^{(d)}$ & 16.8 (1.2) & 0.611 (0.014) & 1.10 (0.04) & 0.45 (0.19) & 5.00 ($<\varepsilon_{2}$) & 11.3 (1.1) & 120 (5) & 162 (7) & 109 	 (6) & 3.35 (0.04) & 0.0 (0.0)\\ 
Sk-69 265 & 1.20 (0.04) & 0.42 (0.05) & 0.68 (0.10) & 0.00 ($<\varepsilon_{2}$) & 5.00 ($<\varepsilon_{2}$) & 56 (5) & 62 (8) & 497 (120) & 4 	 (5)  & 3.454 (0.009) &  56 	 (11) \\ 
Sk-69 270 & 1.19 (0.04) & 0.01 (0.02) & 1.03 (0.28) & 0.00 ($<\varepsilon_{2}$) & 5.00 ($<\varepsilon_{2}$) & 47.2 (2.3) & 131 (15) & 296 (100) & 83 	 (15) & 3.476 (0.010) & 179 	 (11)  \\ 
Sk-69 279 & 0.92 (0.06) & 0.17 (0.09) & 2.3 (0.5) & 0.00 ($<\varepsilon_{2}$) & 5.5 (2.4) & 39 (11) & 52 (22) & 630 (160) & 13 	 (14) &  3.488 (0.012) & 48 	 (13) \\ 
Sk-69 280 & 0.91 (0.04) & 0.05 (0.04) & 1.0 (0.5) & 0.00 ($<\varepsilon_{2}$) & 13.0 (2.3) & 40.7 (2.9) & 81 (8) & 425 (50) & 41 	 (9) & 3.493 (0.012) & 125 	 (13)    \\ 
Sk-70 116 & 1.124 (0.017) & 0.51 (0.03) & 0.95 (0.17) & 0.00 ($<\varepsilon_{2}$) & 5.00 ($<\varepsilon_{2}$) & 25.2 (1.8) & 71.3 (2.8) & 692 (160) & 46.1 	 (2.8) & 3.464 (0.007) &  61.1 	 (6.7)  \\ 
\emph{LMC2 Ss} & 1.09 	 (0.19) & 	0.21 	 (0.14) & 	2.2 	 (1.6) & 	0.00 	 ($<\varepsilon_{2}$) & 	8 	 (4) & 	52 	 (8)  & 	149 	 (60)  & 	415 	 (80) & 	
98 	 (60)  &   3.45 	 (0.04) & \textemdash \\
\\ \hline \hline
 \end{tabular}
\begin{flushleft} 
We adopt $\varepsilon_{1}=0.000001$, $\varepsilon_{2}=0.01$, $\varepsilon_{3}=0.0001$ \\
$^{(a)}$ Fit solution for AzV23 without PAHs \\
$^{(b)}$ Fit solution for AzV23 including PAHs \\
$^{(c)}$ Fit solution for Sk\textendash 69~256 with similar contributions by classical dust and PAHs \\
$^{(d)}$ Fit solution for Sk\textendash 69~256 without PAHs
\end{flushleft}
\end{scriptsize}
\label{classictable}
\end{sidewaystable*}
}

\begin{table*}
\centering
\caption{Mean properties of the PAH mixture (in italic the average ISECs).} 
\label{pahtable}
\begin{tabular}{cccc}
\\
\hline
\hline
\\
\multicolumn{1}{c}{LoS} & \multicolumn{1}{c}{$N_{\rm C}^{\rm {PAH}}/A_V$}  &
\multicolumn{1}{c}{$\langle Q
\rangle/[{\rm C}]$}& 
\multicolumn{1}{c}{ $\sigma_Q /[{\rm C}]$} \\
& \multicolumn{1}{c}{ 10$^{17}$ (cm$^{-2}$ mag$^{-1}$)} & 
\multicolumn{1}{c}{ ($e^-$)} & \multicolumn{1}{c}{ $(e^-)$} \\
\\
\hline \hline
\multicolumn{4}{c}{\footnotesize SMC bar sample} \\
AzV18  & 2.8 	 (0.4)  &   -0.053 	 (0.012) &	0.051 	 (0.013)  \\ 
AzV214    & 	1.9 	 (0.4)   &   -0.007 	 (0.021)  &	0.073 	 (0.010)  \\ 	
AzV23$^{(a)}$ &   3.72 	 (0.21)  &  0.007 	 (0.012) &	0.075 	 (0.004) \\
AzV398 &	2.6 	 (0.3)  &  -0.010 	 (0.017)  & 	0.068 	 (0.004)   \\ 	  
\emph{SMC Bar} & 3.4 	 (0.3)    &   0.006 	 (0.022) &	0.072 	 (0.010)   \\
\hline  
\multicolumn{4}{c}{\footnotesize SMC wing sample} \\
AzV456          & 4.40          (0.19) &        0.060   (0.008) &        0.045          (0.008) \\
\hline
\hline
\multicolumn{4}{c}{\footnotesize LMC average sample} \\
Sk-66 19          & 1.13          (0.17) & 0.015   (0.020) &        0.063          (0.011) \\
Sk-66 88          & 1.06         (0.19) & -0.003  (0.013) &        0.064          (0.006) \\
Sk-67 2          & 3.43          (0.04) & 0.025   (0.001) &        0.051          (0.001) \\
Sk-68 129  & 1.6          (0.3) & -0.012  (0.011) &        0.071          (0.005) \\
Sk-68 23          & 1.40          (0.08) & -0.001  (0.009) &        0.063          (0.005) \\ 
Sk-68 26          & 1.38          (0.10) & 0.000  (0.010) &        0.056          (0.006) \\  
Sk-69 108 & 1.72          (0.22) & 0.004   (0.016) &        0.055          (0.009) \\ 
Sk-69 206 & 0.75          (0.09) & 0.015   (0.008) &        0.056          (0.007)  \\ 
Sk-69 210 & 1.36          (0.09) & 0.026   (0.006) &        0.050          (0.006)  \\
Sk-69 213 & 2.43          (0.26) & 0.034   (0.010) &        0.065          (0.006) \\ 
\emph{LMC} &  2.6 	 (0.6)    &  0.019 	 (0.022)  & 	0.068 	 (0.010)  \\
\hline
\multicolumn{4}{c}{\footnotesize LMC2 Ss sample} \\
Sk-68 140  & 2.18          (0.17) & -0.030  (0.010) &         0.066          (0.007) \\ 
Sk-68 155  & 1.43          (0.16) & -0.009  (0.013) &        0.076          (0.005) \\
Sk-69 228 & 0.53          (0.06) & -0.013  (0.021) &        0.069          (0.012) \\ 
Sk-69 256$^{(b)}$ & 9.5 (1.8)   &     0.05   (0.06)   &   0.07      (0.03)  \\
Sk-69 265 & 1.8          (0.4) & -0.005  (0.007) &        0.061          (0.007) \\ 
Sk-69 270 & 4.06          (0.26) & 0.007   (0.009) &        0.073          (0.007) \\
Sk-69 279 & 0.64          (0.17) & 0.010   (0.018) &        0.065          (0.006) \\ 
Sk-69 280 & 2.24          (0.23) & -0.003  (0.010) &        0.055          (0.007) \\ 
Sk-70 116 & 0.94          (0.10) & -0.006  (0.016) &        0.074          (0.007) \\ 
\emph{LMC2 Ss}  & 2.7 	 (0.6)  &  0.008 	 (0.020)  & 	0.068 	 (0.010) \\
\hline \hline
\end{tabular}
\begin{flushleft} 
We adopt $\varepsilon_{1}=0.001$, $\varepsilon_{2}=0.1$ \\
$^{(a)}$ Fit solution for AzV23 including PAHs \\
$^{(b)}$ Fit solution for Sk\textendash 69~256 with similar contributions by classical dust and PAHs \\
\end{flushleft}
\end{table*}

\begin{table*}
\centering
\caption{Total Si and C abundances for the lines of sight plotted in Figs.~\ref{smcfitcurves} and \ref{lmcfitcurves}. All the abundances are espressed in ppM.}
\label{masstable}
\begin{tabular}{lccccccccc}
\\
\hline 
\hline
\noalign{\vskip0.5mm}
\\
\multirow{2}{1.6cm}{LoS}  & Si/H (ppM)         & &\multicolumn{3}{c}{C/H (ppM)} & Si/C \\
 & & & mantle & PAH & total &  \\
 \\                                  
\hline \hline
\multicolumn{7}{c}{\footnotesize SMC bar sample} \\
AzV18 & 4.8 (1.0)            & &     2.7 (1.3)               & 12.1 (1.6)           & 14.8 (1.6) & 0.3 	 (0.1)  \\
AzV214 &       7.6 (1.2)     && 2.0 (0.4)                  &  11.3 (2.2)           & 13.3 (2.5) & 0.6 	 (0.2) \\
AzV23 $^{(a)}$ &   6.8 ($<\varepsilon$)     & & 65.0  (0.3)        & \textemdash          & 65.0  (0.3) & 0.3 	 (0.1) \\
AzV23 $^{(b)}$          &     6.9 (1.4)       &&        0.7 (1.2)           &        20.0 (1.1) &   20.7 (1.2) &  0.1 	 ($<\varepsilon$) \\
AzV398        &   7.1 (1.7)          &&        3.5 (1.5)           &        14.9 (1.8) &        18.4 (2.4) & 0.4 	 (0.1) \\
\hline
\multicolumn{7}{c}{\footnotesize SMC wing sample} \\
AzV456        &  17.8 (1.1)          & & 9 (5)                     & 93 (4)              & 102 (7) & 0.2 	 ($<\varepsilon$) \\
\hline
\hline
\multicolumn{7}{c}{\footnotesize LMC Average sample} \\
Sk-66 19     & 13.8 (0.6)            & & 4.0 (1.5)                 &  13.1 (1.9)           & 17.1 (2.9) & 0.8 	 (0.1)\\
Sk-66 88      & 20.5 (0.8)           & & 10.7 (1.2)                & 19 (3)                &29 (4) & 0.7 	 (0.1) \\
Sk-67 2       & 18.1 (0.7)           & & 5.4 (0.2)                  & 69.4 (0.9) & 74.8 (1.1) & 0.2 	 ($<\varepsilon$)  \\
Sk-68 129     & 12.0 (1.4)           & & 4.0 (2.3)                  & 16 (3) & 20 (5) & 0.6 	 (0.2) \\
Sk-68 23     &  61.5 (0.8)           & & 14 (4)                 & 78 (4) &    91 (8) & 0.7 	 (0.1) \\
Sk-68 26     & 18.1 (0.6)            & & 10 (4)              & 22.8 (1.7) & 33 (5) & 0.6 	 (0.1) \\
Sk-69 108    &  31.0 (1.1)           & & 9.2 (2.1)               & 50 (6) &59 (8) & 0.5 	 (0.1) \\
Sk-69 206    &  12.0 (0.3)           & & 5.0 (0.8) & 7.9 (1.0)   & 12.9 (1.3) & 0.9 	 (0.1)  \\
Sk-69 210    & 13.5 (0.5)            & & 4.7 (0.7) & 17.6 (1.1)  & 22.3 (1.0) & 0.6 	 ($<\varepsilon$) \\
Sk-69 213    &  17.4 (1.9)           & & 5.3 (1.3) & 43 (5)      & 48 (4) & 0.4 	 (0.1) \\
\hline
\multicolumn{7}{c}{\footnotesize LMC2 Ss sample} \\
Sk-68 140     &  11.5 (1.3)          & & 7.8 (2.7)                  & 25.3 (2.0)  & 33 (4) &0.4 	 (0.1) \\
Sk-68 155     &  17.2 (2.4)          & & 3.9 (0.7)                  & 18.4 (2.0) & 22.2 (1.9) & 0.8 	 (0.1) \\
Sk-69 228    &  16.1 (2.2)           & & 12.7(1.1) & 6.5 (0.7)   & 19.3 (1.5) & 0.9 	 (0.2) \\ 
Sk-69 256$^{(c)}$ & 5.0 (1.0)        && 18 (10)     & 39 (7)      &  57 (13) & 0.1 	 ($<\varepsilon$) \\
Sk-69 256$^{(d)}$ & 10.4 (0.4)       && 21(5)      & \textemdash &  21 (5) & 0.5 	 (0.2)  \\
Sk-69 265    &  10.0 (1.4)           & & 2.0 (0.3) & 12.1 (2.4)  & 14.2 (2.4) & 0.7 	 (0.2) \\
Sk-69 270    &  13.2 (2.7)           & & 3.9 (1.1)  & 50 (3)     & 53.9 (2.9) & 0.3 	 (0.1) \\
Sk-69 279    &  23 (4)               & & 8.7 (1.9)  & 10.4 (2.8) & 19 (3) & 1.2 	 (0.3)\\
Sk-69 280    &  14.7 (1.1)           & & 2.0 (1.2) & 30 (3)      & 32 (3) & 0.5 	 (0.1)\\
Sk-70 116    &  23.3 (1.7)           & & 5.9 (1.0)  & 19.0 (2.1) & 25.0 (2.2) & 0.9 	 (0.1)  \\
\hline \hline
\end{tabular}
\begin{flushleft} 
We adopt $\varepsilon=0.1$ \\
$^{(a)}$ Fit solution for AzV23 obtained with classical dust only (no PAHs)\\
$^{(b)}$ Fit solution for AzV23 including PAHs \\
$^{(c)}$ Fit solution for Sk\textendash 69~256 with similar contributions by
classical dust and PAHs \\
$^{(d)}$ Fit solution for Sk\textendash 69~256 obtained with classical dust only (no PAHs)
\end{flushleft}
\end{table*}

As described in \citet{M13}, the fit is severely underconstrained, mainly due to degeneracy in how many different PAH mixtures can add up to very nearly the same cumulative cross section. As a result, only some of the parameters are well determined by the fit, namely the ones defining the classical dust, and some collective properties of the PAH mixture. Generally, the respective contributions of classic dust and PAHs to the modeled ISECs (as shown in Figs.~\ref{smcfitcurves} and \ref{lmcfitcurves}) are very well constrained. This was the case for all ISECs we fitted in the MWG and for all but two exceptions in the MCs, i.e. AzV23 and Sk\textendash 69~256. In the latter two exceptions, to add insult to injury, there is more than one acceptable solution, with qualitatively different shapes of the classic dust and PAHs contributions that produce almost the same profile. 
\begin{figure}
\begin{center}
 \includegraphics[width=\hsize]{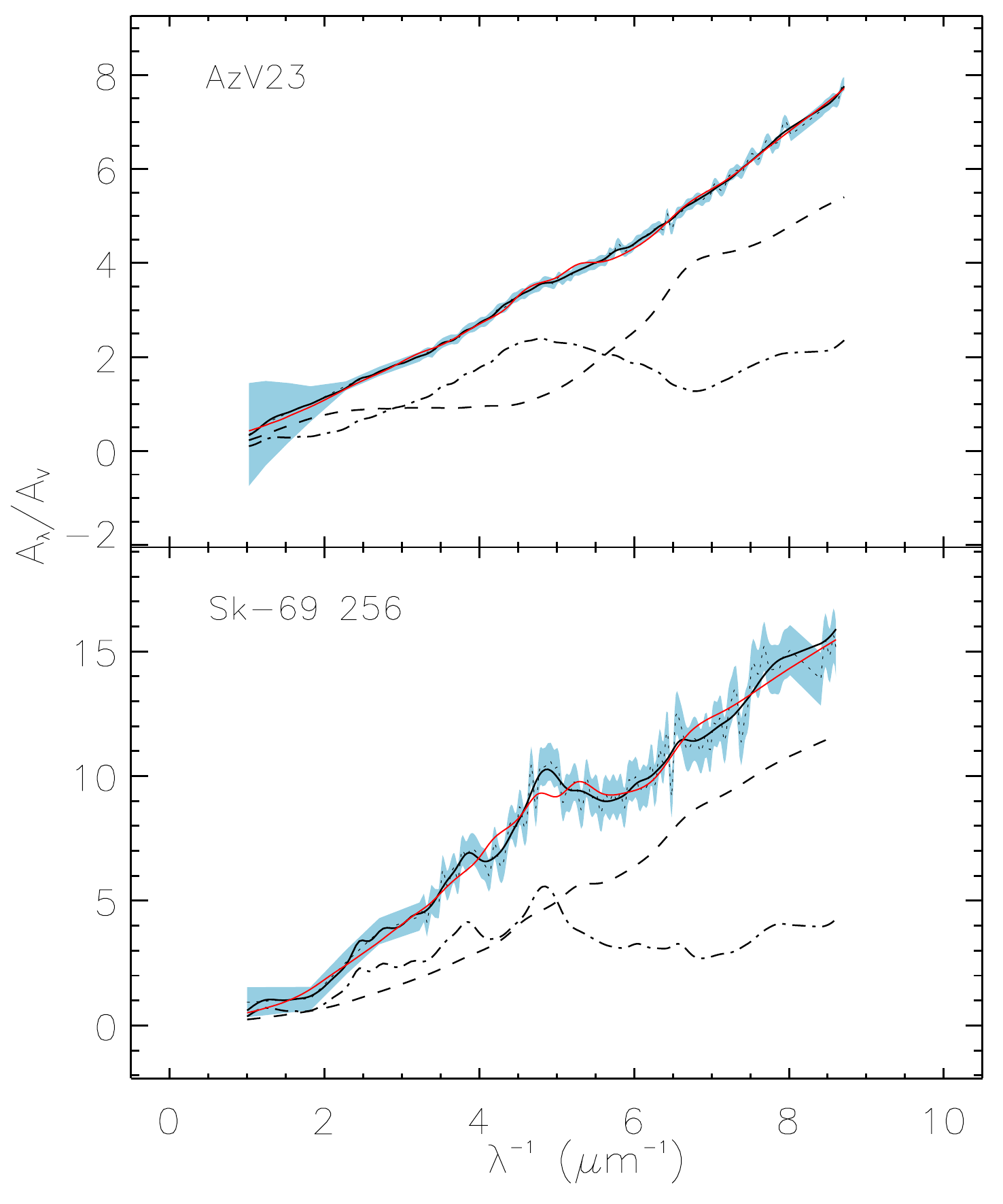} 
\caption{ Alternative (degenerate) acceptable fits for the ISECs towards AzV23 and Sk\textendash 69~256. Black solid lines: absolute best fit. Red solid lines: alternative fit without PAHs. Dashed black lines: classical dust contribution for the best fit. Dot\textendash dashed lines: PAHs for the best fit. The lightly shaded areas are the observational error ranges (see text).}\label{degeneratefits}
\end{center}
\end{figure}
These degenerate fits are shown in Fig.~\ref{degeneratefits}, and parameters for both solutions are given in Tables~\ref{classictable}, \ref{masstable}, and \ref{classicmasstab}.
This reflects an actual physical degeneracy in the possible ways to account for the observed shapes of these specific extinction curves, which could be broken by using additional information (e.g., abundance constraints or observations of aromatic features in emission). AzV23 has an extremely weak bump, if any, and a correspondingly weak non\textendash linear far\textendash UV rise. This can be achieved by a plain monomodal distribution of classic dust with negligible PAHs (red curve in Fig.~\ref{smcfitcurves}); alternatively the same ISEC results from opening a gap in the grain distribution, thereby creating a broad break in the classical dust extinction which almost perfectly cancels with the bump due to PAH absorption. 
\begin{figure}
\begin{center}
 \includegraphics[width=\hsize]{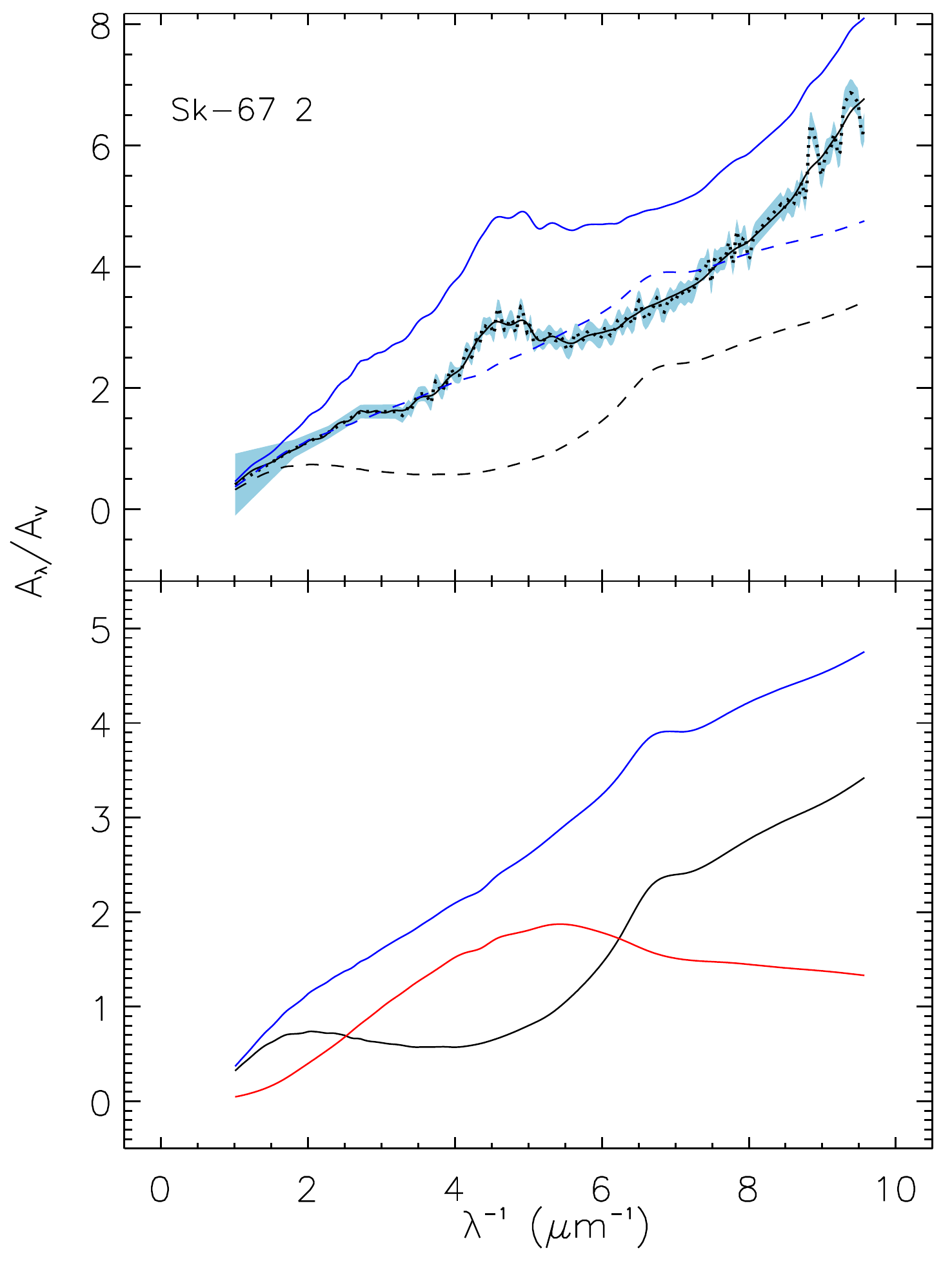} 
\caption{ Effect on the extinction curve of the presence (or absence) of a gap in the size distribution of classical dust grains in the [CM]$^2$ model. Top panel: observed (shaded area) and fitted (solid black curve) ISEC towards Sk\textendash 67~2; the dashed line shows the contribution of classical grains to the extinction. In blue, the total extinction (solid) and the contribution of classical dust grains (dashed line) resulting from exactly the same [CM]$^2$ model with the only difference of artificially removing the gap in the size distribution. Bottom panel: extinction produced by classical dust grains in the best\textendash fitting [CM]$^2$ model (black), in the same model with the size gap artificially removed (blue), and the contribution of dust grains with sizes in the gap (red).}\label{figuranogap}
\end{center}
\end{figure}
To make evident the effect on the extinction curve of a gap in the size distribution of classical dust grains, in the top panel of Fig.~\ref{figuranogap} we show the ISEC towards Sk\textendash 67~2, the best fitting extinction curve (which has a gap) and its components in black, with superimposed, in blue, the extinction curve that would result from exactly the same extinction model but with no gap. 
In the case of Sk\textendash 69~256 there is a hint of a bump, but due also to the low signal\textendash to\textendash noise ratio a similar degeneracy of acceptable solutions occurs, i.e. PAHs may or may not be present. The remaining ISECs in the MC sample do not appear to suffer from such degeneracy, as their fits unambiguously and consistently converge to the same solutions regardless of the starting point of the iterative fitting procedure.

The well\textendash defined PAH collective properties we identify (see \citealt{M13} for details) are the total column density of carbon atoms locked in PAHs per unit visual magnitude, $N_{\rm C}^{\rm PAH}/A_V$ (or given the gas to dust ratio $N_{\rm C}^{\rm PAH}/N_{\rm H}$), and the fractions in each charge state summarized as an average charge per carbon atom, $\langle Q \rangle/[{\rm C}] = \sum_i N_i q_i/ \sum_i N_i {\cal N}^i_{\rm C}$, where $N_i$, $q_i$, and  ${\cal N}^i_{\rm C}$ are column density, charge, and number of carbon atoms for the $i-$th PAH molecule, respectively.  In Table~\ref{pahtable}  we also report the charge dispersion, $\sigma_Q/[{\rm C}]$, a measure of the homogeneity of the charging within the PAH distribution. The total absolute column density of carbon locked in PAHs is not an outcome of the fit, since we are dealing with normalized extinction, but it can derived through the absolute extinction (see Table~\ref{one}). 

The statistics of the MC sample are relatively small (24 lines of sight in total). Nevertheless, some trends are already clear. The fitting results presented in Table~\ref{classictable} occupy a slightly larger volume of the parameter space with respect to the galactic corresponding values. In particular, while for the MWG lines of sight the size distribution of classical grains is always bimodal with no exceptions (see Table~1 in \citealt{M13}), with a gap in size $\Delta a = b_-  - a_+$ between $ \sim 30$ and 200~nm, there are a few MC lines of sight for which the size distribution shows a negligible gap (i.e. AzV214, Sk-69~228 and one of the degenerate solutions for AzV23). In general, while there is considerable scatter in both, median values of $\Delta a$ are smaller in the MCs with respect to those we previously found for the MWG. The same can be seen if one compares the fits for the average MWG ISEC to the average ISECS for the MCs.

Another important difference between the two samples resides in the predominant presence of $sp^3$ carbon mantles in the MCs. The ${sp^2}$ fraction is different from zero only in four cases, namely Sk-68 26, Sk-69~256, Sk-69~210, and AzV23 (see Table~\ref{classictable}), while in the MWG carbonaceous mantles tend to be overwhelmingly aromatic. This is particularly significant as MCs have generally stronger radiation fields than the MWG (up to ten times higher, \citealt{I93}).

From Figs.~\ref{smcfitcurves} and \ref{lmcfitcurves}, it is evident that classical dust dominates the extinction profile, with some exceptions in the bump spectral region. In only one ISEC (Sk-69~256) PAHs may be dominant in the whole range, depending on which of the formally acceptable solutions is the real one.

\begin{figure}
\centering
\includegraphics[width=\hsize]{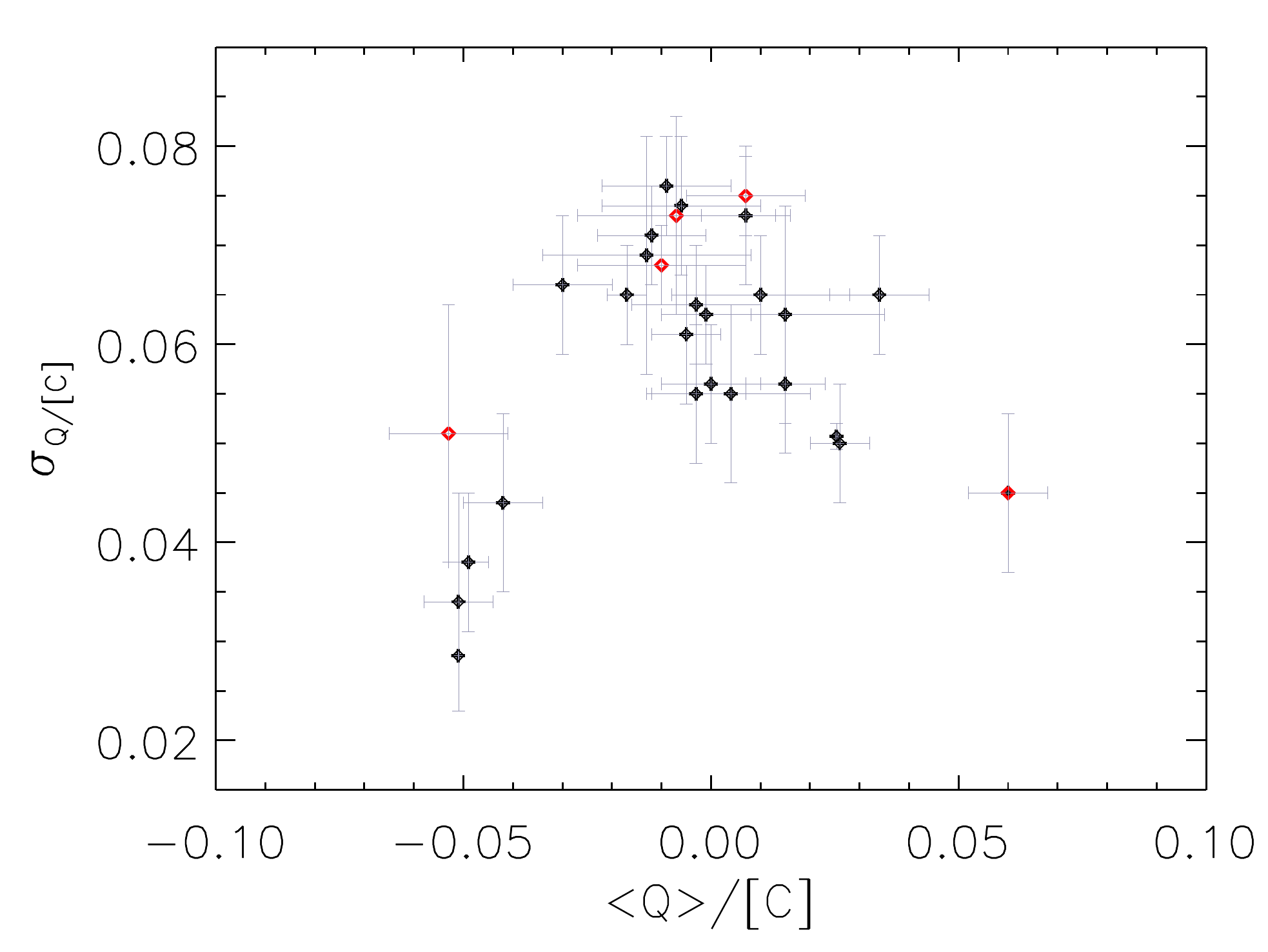} 
\caption{The relation between the mean charge of the PAH collection and its dispersion for each line of sight.}
\label{charge}
\end{figure}

The PAH charge dispersion decreases with increasing PAH charge (Fig.~\ref{charge}) suggesting that when the mixture charge is not negligible all molecular species tend to assume the same charge value, as characteristic during e.g., photoionization (modulo the differences in the photo-absorption cross-sections). On the other side, large charge dispersions are expected when a superposition of different environments, with different physical conditions, are crossed by a line of sight. In this case the average charges of individual regions merge to a median, close to zero, value, but with a large spread around it.

\section{Discussion}\label{discussection}
\begin{figure}
\centering
\includegraphics[width=\hsize]{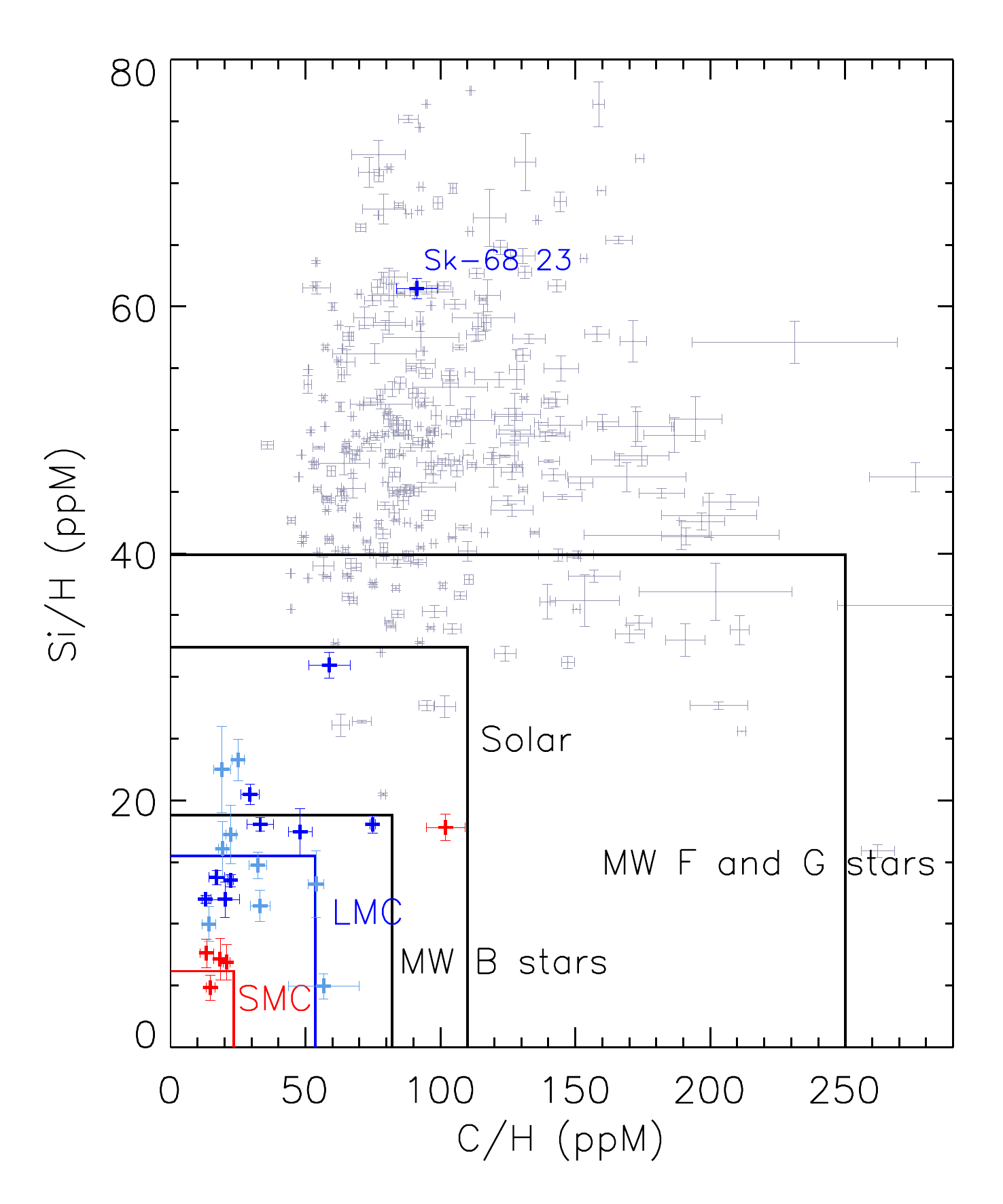} 
\caption{Total abundances of Si and C locked in dust and PAHs in ppM. The circle represents the average MWG ISEC. The boxes refer to observational constraints, values consistent with them being the ones inside each box: solar taken in \citet{Asplund2009}, \citet{Mathis2000}, B, F, and G MWG stars from \citet{Zubko2004}, and SMC and LMC from \citet{Howarth2011}. Gray points are derived for ISECs in the MWG \citep{M13}, dark blue points are ISECs in the LMC ``average'' sample, light blue points in the LMC2 Ss, and red points in the SMC.}
\label{abundances}
\end{figure}
Figure~\ref{abundances} shows in graphical form the total abundances of C and Si locked in dust and PAHs that we obtained in this work for the sightlines to the MCs examined here, also listed in Table~\ref{masstable}, compared with the corresponding values we previously obtained for the MWG \citep{M13} and with some reference values for the total elemental abundances. The level of compatibility of the amounts of Si and C locked in dust in the MCs with abundance constraints is at least as good as that for the MWG, with a similar fraction of data points exceeding mildly the nominal (average) budget of available Si, C constraints almost always respected, one spectacular failure for LMC, one for the SMC and one for the MWG (not visible, because it falls out of the plot). This can be interpreted as a validation of the [CM]$^2$ model for the MCs.
\begin{figure}
\centering
\includegraphics[width=\hsize]{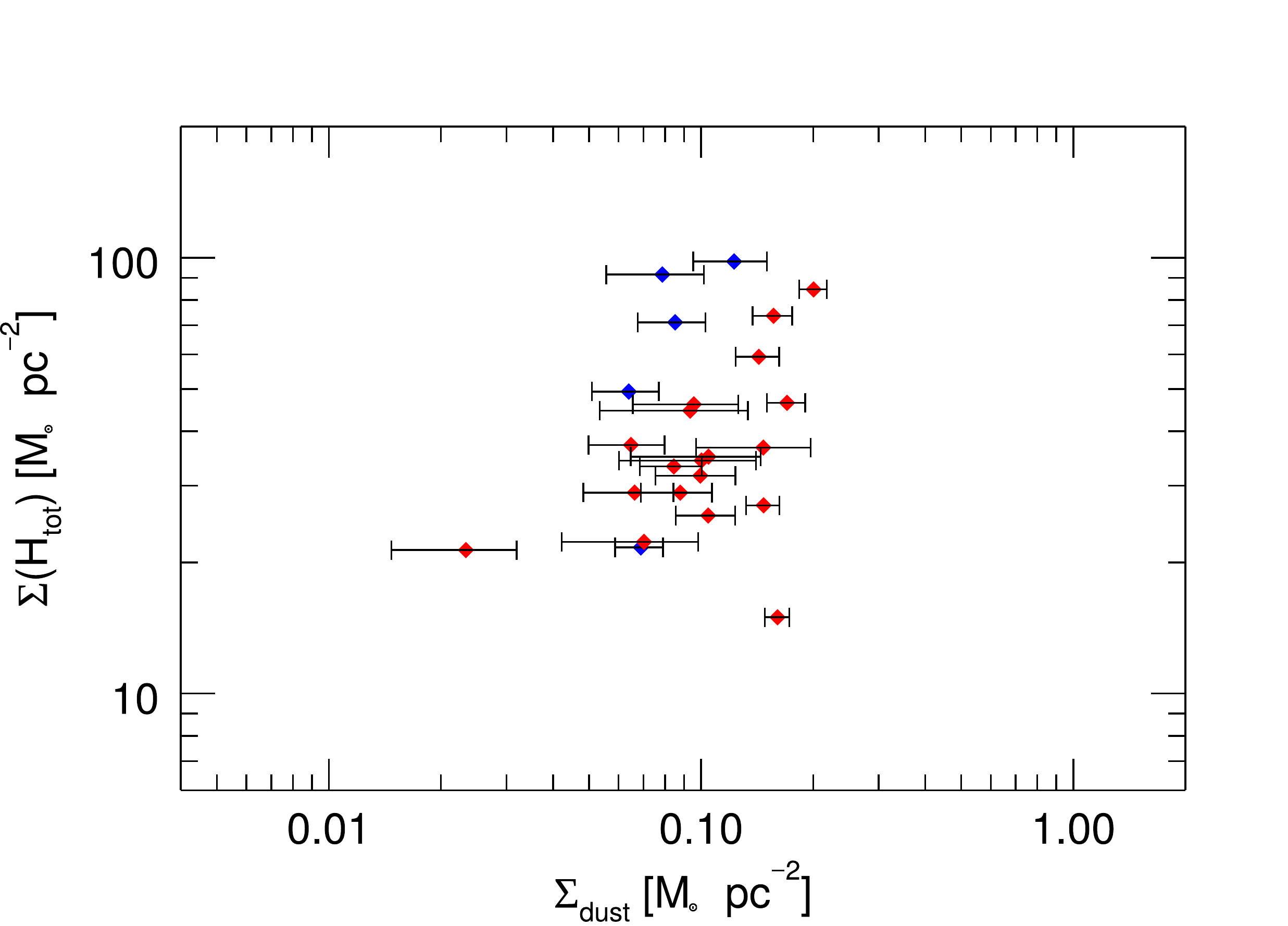} 
\caption{Correlation plot of the mass column densities of classical dust (i.e. excluding PAHs) as inferred from our fits of individual extinction curves, versus the total mass density of H (atomic and molecular, in all charge states) along the same individual lines of sight. Blue dots are SMC lines of sight, red dots LMC ones.}
\label{classicmassfig}
\end{figure}

\begin{table*}
\centering
\caption{Mass column densities of silicates, carbonaceous grain mantles in classical dust, and PAHs, in ${\rm M_{\odot}/pc^2}$, as inferred from our fits of individual extinction curves.}
\label{classicmasstab}
\begin{tabular}{lccccccccc}
\\
\hline 
\hline
\noalign{\vskip0.5mm}
\\
LoS  & $\Sigma(\mathrm{silicates})$ & $\Sigma(\mathrm{sp^2 carbon})$ & $\Sigma(\mathrm{sp^3 carbon})$ & $\Sigma(\mathrm{PAHs})$ \\
\\                                  
\hline \hline
\multicolumn{5}{c}{\footnotesize SMC bar sample} \\

AzV18 & 0.075 (0.023) & 0 (0.0) & 0.0034 (0.0020) & 0.014 (0.004) \\
AzV214 & 0.063 (0.013) & 0 (0.0) & 0.0013 (0.0004) & 0.0068 (0.0016) \\
AzV23$^a$ & 0.083 (0.005) & 0.049 (0.003) & 0.0071 (0.0005) & 0 (0.0) \\
AzV23$^b$ & 0.084 (0.018) & 0 (0.0) & 0.0007 (0.0012) & 0.0180 (0.0015) \\
AzV398 & 0.12 (0.03) & 0.000003 (0.000016) & 0.0047 (0.0021) & 0.018 (0.004) \\
\hline
\multicolumn{5}{c}{\footnotesize SMC wing sample} \\
AzV456 & 0.066 (0.010) & 0 (0.0) & 0.0028 (0.0015) & 0.025 (0.004) \\
\hline
\hline
\multicolumn{5}{c}{\footnotesize LMC Average sample} \\
Sk-66 19 & 0.140 (0.019) & 0.00001 (0.00003) & 0.0033 (0.0013) & 0.0097 (0.0019) \\
Sk-66 88 & 0.163 (0.020) & 0.000001 (0.000004) & 0.0070 (0.0011) & 0.0109 (0.0023) \\
Sk-67 2 & 0.069 (0.028) & 0 (0.0) & 0.0017 (0.0007) & 0.019 (0.008) \\
Sk-68 129 & 0.09 (0.04) & 0.00000 (0.0000011) & 0.0025 (0.0017) & 0.009 (0.004) \\
Sk-68 23 & 0.158 (0.012) & 0.000002 (0.000005) & 0.0028 (0.0008) & 0.0146 (0.0014) \\
Sk-68 26 & 0.095 (0.024) & 0.0019 (0.0016) & 0.0022 (0.0009) & 0.0088 (0.0023) \\
Sk-69 108 & 0.144 (0.015) & 0.00002 (0.00010) & 0.0034 (0.0008) & 0.0169 (0.0028) \\
Sk-69 206 & 0.151 (0.019) & 0.000002 (0.000008) & 0.0052 (0.0010) & 0.0073 (0.0013) \\
Sk-69 210 & 0.196 (0.017) & 0.0038 (0.0008) & 0.0012 (0.0003) & 0.0187 (0.0019) \\
Sk-69 213 & 0.086 (0.019) & 0.000005 (0.000022) & 0.0021 (0.0006) & 0.015 (0.003) \\
\hline
\multicolumn{5}{c}{\footnotesize LMC2 Ss sample} \\
Sk-68 140 & 0.09 (0.03) & 0.00000 (0.0000015) & 0.0050 (0.0024) & 0.015 (0.005) \\
Sk-68 155 & 0.10 (0.04) & 0.0000 (0.0000008) & 0.0019 (0.0007) & 0.0081 (0.0027) \\
Sk-69 228 & 0.09 (0.04) & 0 (0.0) & 0.0061 (0.0027) & 0.0028 (0.0013) \\
Sk-69 256$^c$ & 0.018 (0.008) & 0.0024 (0.0028) & 0.0026 (0.0016) & 0.010 (0.004) \\
Sk-69 256$^d$ & 0.038 (0.014) & 0.0037 (0.0022) & 0.0019 (0.0009) & 0 (0.0) \\
Sk-69 265 & 0.064 (0.015) & 0 (0.0) & 0.00107 (0.00024) & 0.0057 (0.0015) \\
Sk-69 270 & 0.065 (0.018) & 0.00001 (0.00004) & 0.0015 (0.0005) & 0.018 (0.004) \\
Sk-69 279 & 0.14 (0.05) & 0 (0.0) & 0.0045 (0.0017) & 0.0048 (0.0020) \\
Sk-69 280 & 0.084 (0.016) & 0 (0.0) & 0.0009 (0.0006) & 0.0126 (0.0026) \\
Sk-70 116 & 0.102 (0.019) & 0.000001 (0.000003) & 0.0021 (0.0005) & 0.0062 (0.0012) \\
\hline \hline

\end{tabular}
\begin{flushleft} 
$^{(a)}$ Fit solution for AzV23 obtained with classical dust only (no PAHs)\\
$^{(b)}$ Fit solution for AzV23 including PAHs \\
$^{(c)}$ Fit solution for Sk\textendash 69~256 with similar contributions by
classical dust and PAHs \\
$^{(d)}$ Fit solution for Sk\textendash 69~256 obtained with classical dust only (no PAHs)
\end{flushleft}
\end{table*}
Table~\ref{classicmasstab} reports the column mass densities derived from our fits of the individual extinction curves, while Figure~\ref{classicmassfig} shows the correlation of the total dust mass column densities versus total hydrogen mass column density. They can be compared with the results of \citet{RomanDuval2014}, obtained fitting a dust emission model to Herschel data. While of course our data points are very sparse, compared to all the pixels in the maps of the MCs, as used by \citet{RomanDuval2014}, they do fall in the same area of the $\Sigma_\mathrm{dust}$\textendash  $\Sigma(H_\mathrm{tot})$ plane covered by their data, and are therefore fully consistent with them. This is rather interesting, since the two datasets were obtained using completely different observations (emission versus absorption) and dust models.

\begin{figure}
\centering
\includegraphics[width=\hsize]{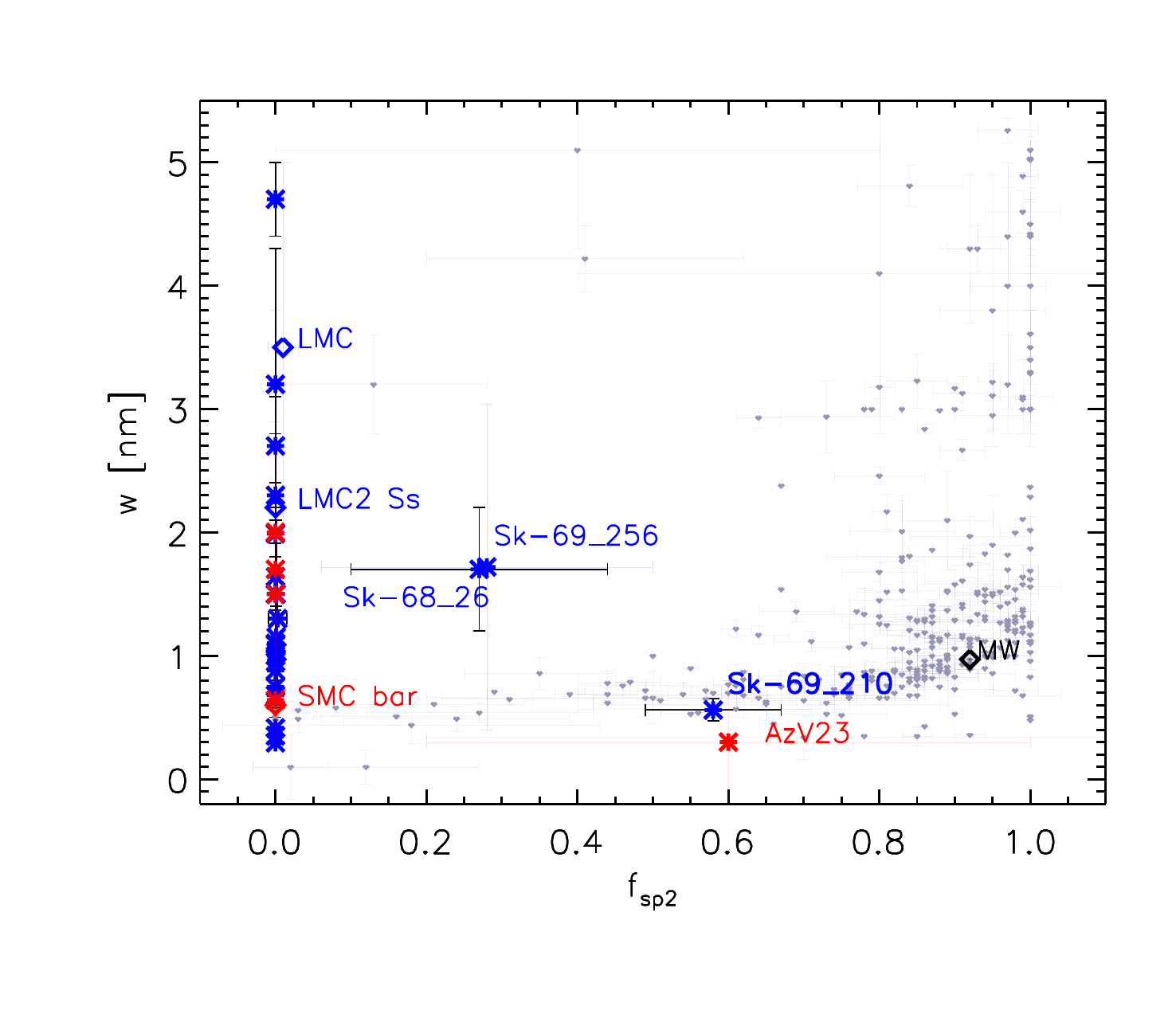} 
\caption{Observationally derived distribution of mantle thickness $w$ and normalized $sp^2$ mantle carbon fraction, with relative errors for the SMC (red asterisks), the LMC (blue
asterisks), and the MWG (pale grey symbols). The black symbol marked ``MW'' represents the values corresponding to the average galactic ISEC. Values corresponding to average ISECs in the MCs are also indicated. MWG data are taken from \citet{M13}.}
\label{MCevofig}
\end{figure}

In \citet{CCP14} we outlined a self\textendash consistent evolutionary scenario for dust in the diffuse ISM (DISM) in the MWG. In such model, silicate grain particles, being relatively long\textendash lived, progressively accrete C atoms in an H\textendash rich environment (the DISM), thereby forming aliphatic carbonaceous mantles. The older layers of these mantles are progressively photoprocessed to aromatic, while new aliphatic material simultaneously continue to deposit on top of them. The deposition of carbon is assumed to be proportional to the rate of collisions of C atoms onto dust grains, with a sticking factor calibrated to match observational constraints on CO abundances. Since dust evolution is modelled in diffuse and translucent clouds, which are rich in atomic H, all freshly deposited carbonaceous material is assumed to be highly hydrogenated, i.~e. aliphatic. Upon UV irradiation, carbonaceous material is assumed to be dehydrogenated, becoming more and more aromatic. To simplify the model, carbonaceous material, instead of having a continuous, smooth transition between deeper, older, and thus more aromatic material, and freshly deposited, aliphatic material, is represented as just two layers, the deeper one completely aromatic, the other completely aliphatic. Occasionally, shocks peel away most of this material, restarting the cycle and ejecting the fragments of carbonaceous mantles into the gas, where they can possibly be a source of PAHs. Shocks are not explicitely included in the set of differential equations giving the time evolution of the grain mantles, but are taken into account in terms of ``instantaneous'' episodes, removing most of the carbonaceous mantles, rehydrogenating the small leftover, and putting back into the gas phase the resulting carbon; the time evolution, according to the differential equations of the model, is then resumed from these new initial conditions.

Upon trying to use this kind of evolutionary scenario to interpret the ISECs observed in the MCs, as shown in Fig.\ref{MCevofig}, it is apparent that the annealing time scale must be almost always much longer than the average time between mantle\textendash shattering events. The annealing scale is defined by the competition between energetic processing by UV radiation (and cosmic rays) and re\textendash hydrogenation by hot H atomic gas, turning aromatic carbonaceous material back into aliphatic. In the MWG, re\textendash hydrogenation was deemed to be by and large negligible, but this may be not the case in the MCs. Carbonaceous mantles in the MCs result from the fit to be almost always completely aliphatic, with only few exceptions. This points to significantly higher star\textendash formation activities, and associated supernova rates, and/or much longer photoprocessing timescales, possibly due to more effective re\textendash hydrogenation. This is in agreement with the current star formation rates in the MCs ($\sim 0.04$ in the SMC, \citealt{B11} and $\sim 0.4$~$M_\odot$~yr$^{-1}$ in the LMC, \citealt{HZ09}) that per unit area are an order of magnitude higher than in the MWG ($\sim 1 \, M_\odot $~yr$^{-1}$, \citealt{RW10}). In general, observations of stellar populations in dwarf galaxies suggest that the star formation bursts are sporadic, separated by millions to billions of years even in isolated systems (e.g., \citealt{W08}). According to their star\textendash formation histories \citep{HZ04,HZ09,R14}, the MCs appear to be typical members of this class of galaxies. MC\textendash type ISECs may be then prototypical of dwarf galaxies of low metallicity, with high gas content (see e.g., \citealt{CCP14b}). 

In the MWG DISM the photo-darkening time is $\tau_{\rm pd} \sim 10^{5} /\chi$~yr \citep{I84,M01}, $\chi$ being local fluctuations around the average intensity of the interstellar UV radiation field. Such time should be compared with lifetime of carbon mantles against sputtering (whose estimate is not a trivial task). Assuming a plausible shock frequency of $\sim 2 \times 10^{-7}$~yr$^{-1}$ \citep{D87}, we derive that dust mantles in the MWG DISM have enough time to be converted in aromatic structure. \citet{CCP14} show that the evolutionary age of ISECs peaks at times of few Myr (assuming typical diffuse interstellar conditions) comparable to the probable average age of diffuse clouds. Translating these considerations to the DISM in the MCs, we may conclude that the carbon recycling time must be much shorter than the local photo\textendash darkening time, $\tau_{\rm pd}  \la 10^4$~yr ($\chi = 10$, \citealt{I93}).  However, such result should be taken with care, as many effects are concurring in dust destruction and formation processes. For instance, small grains are more susceptible to destruction by sputtering in shock waves than large grains \citep{S04}, because larger grains decouple from the gas and are therefore less exposed to ions. In strong shocks, grain\textendash grain interaction may lead to shattering and thus to the creation of smaller grains \citep{A13}, which are then more likely to be sputtered away. Hence, the timescale of dust destruction may be inversely proportional to the abundance of dust, since the rate of collisions is proportional to the grain number density (e.g., \citealt{M14,D14}).

Although PAHs are clearly less abundant in the MCs than in the MWG (see Table~\ref{pahtable}), a potential problem in our description of the carbon cycle in galaxies is posed by the the sharp decline of the  source of aromatic material in the DISM, since most of dust mantles are destroyed before they experience a substantial annealing. PAH formation rates in the cold, carbon\textendash rich winds of evolved stars are too slow, the injection time being approximately 2~Gy. Recent \emph{Spitzer} surveys in SMC also revised down the contribution of AGB stars to dust production (e.g., \citealt{B12}). Without a net positive contribution from supernovae to the dust budget, this suggests that dust must grow in the DISM or be formed by another yet unknown mechanism (see \citealt{Ma11} and \citealt{MK11}). Recent \emph{Herschel} photometric and spectroscopic observations of the supernova 1987A by \citet{Ma15} may suggest that supernovae can be an important source of dust in the interstellar medium, from local to high-redshift galaxies. However, the point is controversial as it depends on the efficiency of supernova remnants in destroyng interstellar grains. \citet{T15} determined dust lifetimes in the MCs exploiting the available data of supernova remnants for these galaxies. These authors inferred dust lifetimes in the MCs significantly shorter then in the MWG, as the cumulative dust production rates by AGB stars and supernovae are one order of magnitude lower than dust destruction in interstellar shocks produced by the expansion of supernova remnants

Conversely, we are left with the puzzle of what becomes of the aliphatic fragments released into the DISM of the MCs by the shattering events. \citet{DW84} proposed that they should be rapidly evaporated into polyyines, and thereafter quickly photodissociated under DISM conditions. They would therefore have a negligible effect on the extinction. If instead a significant population of aliphatic carbonaceous nanoparticles were able to survive for a significant time, thereby achieving non\textendash negligible abundances, they should be visible both in extinction, exhibiting $\sigma^\star \leftarrow \sigma$ features contributing mainly to the non\textendash linear far\textendash UV rise, and given their small sizes possibly producing aliphatic C\textendash H emission. Indeed, some emission in the aliphatic C\textendash H stretch at 3.4~$\mu$m is detected toward the MCs \citep{boulanger2011,mori2012}.

Classical dust extinction is generally larger than PAH extinction with a few exceptions. In these latter cases, the PAH contribution dominates only in the bump region. In all these relatively ``PAH\textendash rich'' cases, the classical size distribution is strongly bimodal, with a gap ranging from 40 to 130 nm, with a tendency to increase the non linearity of the far\textendash UV rise with increasing size breaks. All the remaining ISECs are reproduced with classic grain size distributions that tend to have markedly lower gaps (see Table~\ref{classictable}). 

In \citet{M13} we concluded that, in the framework of [CM]$^2$ model, bumpless extinction curves are not necessarily lines of sight devoid of PAHs, but rather that the $\pi^\star \leftarrow \pi$  PAH absorption can be very effectively masked by a large gap between the maximum size limit of small dust grains and the minimum size limit of large ones. MWG data were not conclusive since MC-type ISECs are rare in our Galaxy. The present results suggest a different interpretation based on the fact that some ISECs show a more or less weak non-linear far-UV rise without showing an appreciable bump. This is particularly evident in the SMC sub sample. In the LMC the most evident trends are small bumps with linear far-UV rises. Within the current model dust populations, for the SMC-type ISEC the only possibility is the introduction of PAHs that, through $\sigma^* \leftarrow \sigma$ absorptions, may produce such rise. However, as a by product, PAH molecules exhibit  $\pi^\star \leftarrow \pi$ transitions who generate the bump. Thus, the only viable solution, for bumpless ISECs, is the removal of mid-size grains, whose extinction power falls in the bump region, from the grain size distribution. 

This may be interpreted in two alternative ways. The first one is that we must be missing an $sp^3-$dominated carbon component, whose sizes are intermediate between small solids and macromolecules. This is also hinted by the numerical tendency during the fitting procedure  to go below the lowest limit of 5 nm in particle size, that is systematic for all the lines of  sight in the MWG and in the MCs. Such small particles might be carbon mantle fragments, possibly produced during some destructive events (as discussed in the previous paragraph), such as e.~g. nanodiamonds \citep{Rai2010,Rai2012,Rai2014}. 

The second interpretation is instead that in our model we are missing no important component contributing to extinction, and hence the fact that PAHs are abundant only if middle\textendash sized dust grains are absent is real, and must have a sound physical explanation. One might argue that if an interstellar cloud, and the dust therein, underwent some shocks in its history, they may have shattered middle\textendash sized grains, producing fragments (small grains) and releasing PAHs from the disrupted carbonaceous mantles. This would be a consistent explanation for the MWG, where carbonaceous mantles are almost entirely aromatic, so that their disruption would indeed produce PAH\textendash like fragments. But in the MCs carbonaceous mantles result to be almost entirely aliphatic, so their destruction can hardly be a viable source of PAHs.

In the case of LMC\textendash type ISECs the results of the fitting procedure are more straightforward than for the SMC\textendash type, bumpless ones: in this galaxy PAHs are present but are a minor component, their presence being only evident in the bump spectral region. The possible extra aliphatic "mesoscopic" component is marginal in this case. Conversely, the problem of the lack of a suitable source of aromatic material is therefore more pressing for LMC than SMC.

\section{Conclusions}

We here concisely summarise the main conclusions of this work.
\begin{itemize}
\item The [CM]$^2$ dust model can very precisely match all the ISECs observed in the MCs, yielding estimates for the total dust mass column density versus total hydrogen column density that are fully consistent with those obtained by \citet{RomanDuval2014} in a completely different, independent way.
\item Abundance of elements locked up in [CM]$^2$ models fitting the ISECs in the MCs are at least as compatible with abundance constraints as those for the MWG \citep{M13}.
\item When ISECs have vanishingly weak bump and far\textendash UV non\textendash linear rise, they can be equally well fitted either by classical dust alone or by classical dust \emph{and} PAHs. In such cases models without PAHs use much more Si and C atoms, which may put them at odds with abundance constraints.
\item Carbonaceous mantles in the MCs resulting from the [CM]$^2$ model fits are overwhelmingly aliphatic, whereas those in the MWG are predominantly aromatic. Together with the observational constraint that radiation fields in MCs are typically 10 times stronger than the average in the MWG \citep{I93}, in the [CM]$^2$ model this implies that mantle\textendash shattering events must be $\gg 10$ times more frequent in the MCs than in the MWG, and/or much more efficient. Desorption from such mantles cannot be a significant source of PAHs in the MCs.
\item Bumpless ISECs with large non\textendash linear far\textendash UV rises can be fitted within the [CM]$^2$ model, but this requires a finely tuned match of the gap in the size distribution of classical dust grains, resulting in a dip in the extinction they produce that precisely cancels with the bump. Either there is some hitherto not understood physical relation connecting the population of PAHs with the size distribution of silicate grains in the diffuse interstellar medium, or this hints that the [CM]$^2$ model must be missing some component different from PAHs that can can produce a non\textendash linear far\textendash UV rise without simultaneously contributing to the bump.
\end{itemize}

\section*{Acknowledgments}
We thank the referee for her/his constructive criticism, that helped us to improve this work.
We acknowledge the support of the Autonomous Region of Sardinia, Project CRP 26666 (Regional Law 7/2007, Call 2010)

\end{document}